\documentclass[aps,twocolumn,prl]{revtex4}
\usepackage{amsmath}
\usepackage{graphicx}

\begin{document}

\title{Band structure approach to the resonant x-ray scattering}
\author {I.S.~Elfimov$^1$, N.~Skorikov$^2$, V.I.~Anisimov$^2$ and
         G.A.~Sawatzky$^1$}
\affiliation{
$^1$Laboratory of Applied and Solid State Physics, Materials Science Center,
University of Groningen, Nijenborgh 4, 9747 AG Groningen, The Netherlands\\
$^2$Institute of Metal Physics, Russian Academy of Sciences,
620219 Ekaterinburg GSP-170, Russia\\}

\begin{abstract}
We study the resonance behaviour of the forbidden 600 and 222
x-ray Bragg peaks in Ge using LDA band structure methods. These
Bragg peaks remain forbidden in the resonant dipole scattering
approximation even taking into account the non local nature of the
band states. However they become allowed at resonance if the
eigenstates of the unoccupied conduction band involve a
hybridization of $p$ like and $d$ like atomic states. We show that
the energy dependence of the resonant behaviour, including the
phase of the scattering, is a direct measure of this $p$-$d$
hybridization.and obtain quantitative agreement with experiment. A
simple physical picture involving a product of dipole and
quadrupolar transition matrix elements explains this behaviour and
shows that it should be generally true for cases where the
resonating atom is not at an inversion center. This has strong
implications for the description of the resonance behavior of
x-ray scattering in materials where the resonant atom is not at an
inversion center such as  V$_2$O$_3$ and in ferro and antiferro
electric and piezo electric materials in general.
\end{abstract}


\maketitle

In a recent paper T.L.~Lee {\it et al} \cite{Colella01} describe a
detailed study of the energy dependence of the 600 reflection in
the vicinity of the Ge K-edge. They demonstrate a strong 600
reflection close to the Ge K-edge with an energy dependence quite
different from the absorption edge structure in that the 600
intensity drops of rapidly at energies just beyond the edge. The
intensity of this reflection at the edge corresponds to a
structure factor of 0.45 electrons. The fact that one sees this
reflection at all as well as the large intensity and the peculiar
energy dependence is difficult to explain. This because, even if
we take into account the non spherical nature of the charge
distribution and put electron charge in the bonds between the
atoms from which the x-rays, scatter the 600 reflection remains
forbidden. Also the usual dipole-dipole  or quadrupole-quadrupole
like resonances as discussed for the manganites \cite{Murakami98,
Ishihara98, Elfimov99, Benfatto99} and V$_2$O$_3$
\cite{Paolasini99, Ezhov99, Mila00} cannot contribute to this
reflection as discussed below. Recently Templeton and Templeton
\cite{Templeton94} suggested that the lowest order process which
will allow such a reflection is one involving a product of a
dipole and a quadrupole like matrix element for the scattering. In
this paper we show that band structure calculations provide a
natural and detailed explanation for the observed behaviour and
that the resonance behaviour is a direct  measure of the energy
dependence of the $p$-$d$ hybridization in the conduction band of
Ge. This is, in a way equivalent to Templeton's suggestion but
here is caused by a natural $p$-$d$ mixing due to a lack of
inversion symmetry at the Ge atomic sites.

We also show that the 222 reflection becomes allowed in this
fashion but this is not as spectacular since the 222 reflection is
also weakly allowed by simple Thomson scattering of the charge
situated in the bonds between the atoms \cite{Ewald36, Colella66,
Colella74}. Non the less the energy dependence of the phase of
this extra scattering channel causes a decrease in the net
scattering close to the edge as observed and we predict an
increase at energies about 30 eV above the edge because of a
change in the phase of the $p$-$d$ mixing at these energies.
Although the case of Ge may, for some, not be that interesting it
serves as a demonstration of the effect and the success of theory
to explain it. We suggest that this effect will play an important
role in the resonant behaviour of recently discussed controversial
materials like V$_2$O$_3$, Fe$_2$O$_3$ and many more vanadates ,
titanates and manganites which often also lack inversion symmetry
at the resonating ion. In fact quite generally this effect could
be very important in ferro and antiferro electric as well as
piezoelectric materials. This may also provide a simple way to
actually distinguish between materials with and without an
inversion center which is an important problem in crystallography.

\begin{figure}
\centering
\includegraphics[clip=true,width=0.23\textwidth]{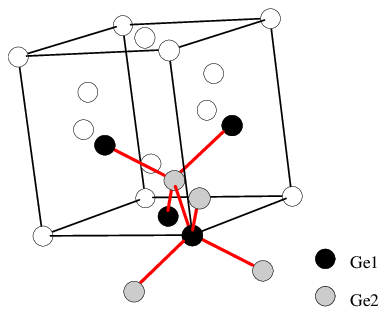}
\caption{The diamond crystal structure.} \label{fig:str}
\end{figure}

Ge has the diamond structure which contains 8 atoms per unit cell
as depicted in Fig.\ref{fig:str} showing also the location of
these atoms in a simple cubic notation. We notice immediately that
if all the Ge atoms are equivalent then reflections like the 600
and 222 reflections will be forbidden. Even if we take into
account that the charge density around each atom is not spherical
but somewhat enhanced along the bond directions we still will find
zero intensity for the 600 reflection because of the space group
symmetry. However in this case the 222 reflection can become
weakly allowed in a purely Thomson scattering mechanism. Looking
at Fig.\ref{fig:str} more carefully however and at the Ge atoms
connected by a bond it is easy to see that in fact Ge$_1$ and
Ge$_2$ are not really equivalent in that the charge distribution
around atom $1$ is the, by 180 degrees, rotated image of that
around atom $2$. In Thomson scattering this difference  will not
be noticed however. This difference between Ge$_1$ and Ge$_2$ is
caused by a change in sign of the hybridization between $s$ and
$p$ and the $p$ and $d$ wave functions forming the eigenstates of
the occupied valence and unoccupied conduction band states. For
example if we take the Ge$_1$-Ge$_2$ direction to be the z
direction then a wave function like $\alpha s+\beta p_z+\gamma
d_{3z^2-r^2}$ on Ge$_1$ would have charge pile up in the
Ge$_1$-Ge$_2$ bond but for Ge$_2$ this combination at the same
energy would be $\alpha s-\beta p_z+\gamma d_{3z^2-r^2}$. Notice
the sign change of the $p$ contribution because of the ungerade
character of its wave function.We note that $s$, $p$, $d$ can mix
in this structure because of the lack of inversion symmetry about
the Ge atoms.

We will not see this difference in an experiment which is
sensitive only to the charge density as in Thomson scattering.
However in a resonant scattering process the anomalous form factor
is given by
\begin{equation}
A \sim \sum_b \frac{<a|\vec{\epsilon}\,'\cdot\vec{r}
                       \,\,\,\,e^{-i\vec{K}'\cdot\vec{r}}|b>
                    <b|\vec{\epsilon}\cdot\vec{r}
                       \,\,\,\,e^{i\vec{K}\cdot\vec{r}}|a>}
                   {E_a-E_b+\hbar\omega-i\Gamma/2}
\label{eq:basic}
\end{equation}
Where $\epsilon'$ and $\epsilon$ are the polarization of the
scattered and incident radiation respectively and $K'$ and $K$ the
respective wave vectors. For the K-edge studies $a$ refers to a
1$s$ orbital and $b$ to a band eigenstate.

Following Templeton we expand the exponential as
$1+i\vec{K}\cdot\vec{r}$ resulting in 3 kinds of contributions:
dipole-dipole, quadrupole-quadrupole and the cross terms
dipole-quadrupole.  The dipole-dipole and quadrupole-quadrupole
terms will be the same for atoms $1$ and $2$ because they involve
only the squares of the wave functions and therefore the sign
difference in the $p$ component for atoms $1$ and $2$ disappears.
Therefore we concentrate only on the dipole-quadrupole cross terms
which for each of the two atoms is given by:
\begin{multline}
A_{dq} \sim i \sum_b
              \bigl ( \frac{<a|\vec{\epsilon}\,'\cdot\vec{r}
                               \,\,\,\,\vec{K}'\cdot\vec{r}|b>
                            <b|\vec{\epsilon}\cdot\vec{r}|a>}
                           {E_a-E_b+\hbar\omega-i\Gamma/2}-\\
                      \frac{<a|\vec{\epsilon}\,'\cdot\vec{r}|b>
                            <b|\vec{\epsilon}\cdot\vec{r}
                              \quad \vec{K}\cdot\vec{r}|a>}
                           {E_a-E_b+\hbar\omega-i\Gamma/2} \bigr )
                           \nonumber
\end{multline}
\begin{figure}
\centering
\includegraphics[clip=true,width=0.43\textwidth]{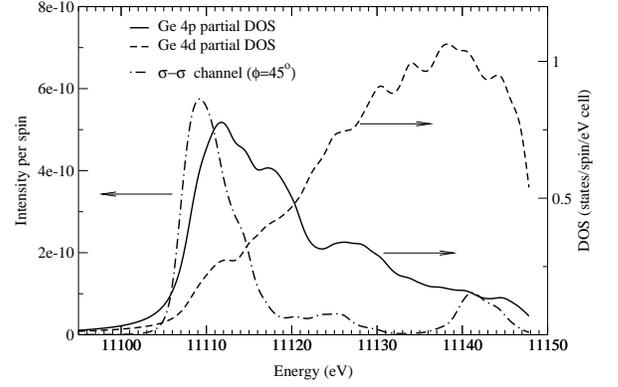}
\caption{Ge partial density of states and the intensity of 600
reflection. The DOS and structure factor are broadened by 3 eV
Lorentzian.} \label{fig:bpdos}
\end{figure}

The two terms are obviously not equal yielding a net result for
each. For the K-edge we see that the $r^2$ terms get us from the
ground state to the $d$ component of the excited state
eigenfunction and the dipole ($r$) term gets us back from this
eigenstate via the $p$ component to the ground stat or visa versa.
Immediately we see that the phase of the $p$-$d$ mixing in the
eigenstate enters and therefore the two Ge atoms discussed above
will in fact scatter differently. Now one might say that this
would also be the case for the $s$-$p$ hybridization but a $s$-$s$
transition is not quadrupole allowed because an $s$ orbital has no
quadrupole moment. This kind of hybridization could only be
observed in a combination of a dipole and monopole transition.

Now that we understand which physical scattering mechanism can
distinguish between these two Ge atoms we can check if there is
such a difference seen in the electron structure of Ge. To do this
we look at the conduction band part of the band structure in a LDA
calculation (TBLMTO-47 computer code \cite{lmto}), where we
neglect the core hole potential. The 4$p$ projected local density
of states would correspond to the energy dependence of the K-edge
x-ray absorption spectrum broadened by the core hole life time.
This core hole life time is about 3 eV. The 600 and 222 reflection
structure factors in this mechanism will be determined by the
energy dependence of the product of the amplitude of the $p$ and
$d$ component of the band eigenstate times a radial integral and
is given by :
\begin{equation} \label{eq:amp_in_details}
\begin{split}
A_{dq}\sim i\sum_n e^{-i\vec{G}\cdot\vec{R_n}} \sum_{m=-l}^{l}
\sum_{m'=-l'}^{l'}
&\hat{\epsilon}\,'\,\bar{d}_{m}\,\hat{K}'\,\,T_{lm,\,l'm'}^n
\,\,\hat{\epsilon}\,\vec{p}_{m'}-\\
&\hat{\epsilon}\,'\,\vec{p}_{m'}\,\,T_{l'm',\,lm}^n\,\,
\hat{\epsilon}\,\bar{d}_{m}\,\hat{K} \nonumber
\end{split}
\end{equation}
where $\vec G = \vec K' - \vec K$, $m$ and $m'$ are magnetic
quantum number of $d$ and $p$ orbital respectively
\begin{equation}\label{eq:T_tensor}
\begin{split}
T_{lm,\,l'm'} =& \sum_j \int d\vec{k}
                  \frac{C^j_{lm}(\vec{k})C^{j *}_{l'm'}(\vec{k})}
                       {E_{1s}-E_{j}(\vec{k})+\hbar\omega-i\Gamma/2} \\
p^i_{m} =& \int d r R_{1 s}(r) r^3 R_{4 p}(r)
            \int d\Omega \,Y_{0 0} \hat{r}^i Y_{1 m} \\
d^{ij}_{m} =& \int d r R_{1 s}(r) r^4 R_{4 d}(r)
               \int d\Omega \,Y_{0 0} \hat{r}^i \hat{r}^j Y_{2 m} \nonumber
\end{split}
\end{equation}
j is a band index, $C^j_{m}(\vec{k})$ are components of the
eigenvector projected on a particular orbital of site $n$ with
lattice vector $\vec R_n$ and $Y_{lm}$ are cubic spherical
harmonics.

\begin{figure}
\centering
\includegraphics[clip=true,width=0.23\textwidth]{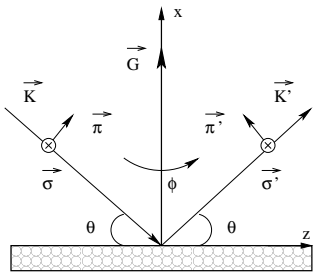}
\caption{The geometry of x-ray scattering.}
\label{fig:geom}
\end{figure}

In Fig.\ref{fig:bpdos} we show the 4$p$ projected partial density
of states and this convoluted with a Lorentzian of width 3eV. We
note that we have not taken into account higher lying $p$ bands
than the 4$p$ band to which dipole transitions are of course also
allowed so the experimental absorption edge will fall off somewhat
less rapidly than calculated here. To calculate the resonant
amplitude we use the scattering geometry of Fig.\ref{fig:geom} and
the angles determined by the Bragg conditions for the given
energies. At a photon energy equal to 11107 eV the  scattering
angle is 36.35 degrees for the 600 reflection and 20.01 degrees
for the 222 reflection. The resonant intensity will  be determined
by the product of the $p$ and $d$ amplitude at a given energy in
the band structure integrated over k space and summed over atoms
$1$ and $2$ with the appropriate phase factor
$e^{-i\vec{G}\cdot\vec{R}_n}$. As discussed above the product of
the $p$ and $d$ amplitude for atoms $1$ and $2$ will have opposite
sign which in the sum will be cancelled by the opposite phase
factor for the reciprocal wave vector 600 or 222.

In Figure \ref{fig:bpdos} we also show the 4$d$ partial density of
states and we see that in general the 4$d$ states are as expected
higher in energy than the 4$p$ states. It is then easy to see that
there will be a phase change in the $p$-$d$ mixing as we go from
the edge to the high energy states at about 30 eV higher. The
doted dashed curve is then the resonant scattering intensity
calculated for the scattering geometry of Fig.\ref{fig:geom} which
demonstrates the strong energy dependence of the $p$-$d$ mixing
resulting in a strong peak close to threshold and a weaker
structure 16 to 18 eV higher in energy and another peak at 30 eV
above threshold. In fact the energy scans done to date did not
cover the high energy region so this should be considered as a
prediction which should be checked experimentally. The first
strong peak at threshold agrees very well in intensity and width
with the experimental data of Lee {\it et al} \cite{Colella01}.

\begin{figure}
\centering
\includegraphics[clip=true,width=0.43\textwidth]{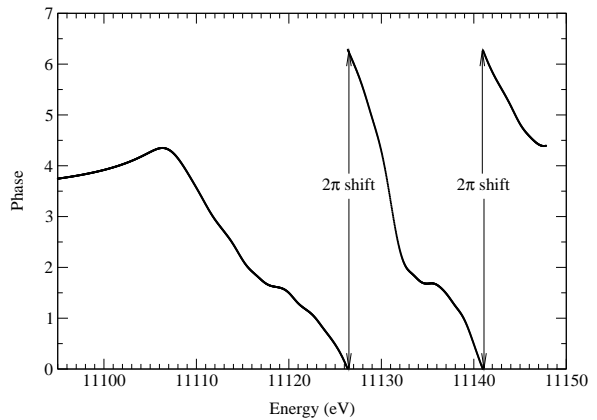}
\caption{The energy dependence of the phase of the 600 reflection
in $\sigma$-$\sigma$ ($\pi$-$\pi$) channel with $0<\phi<90$ or
$180<\phi<270$; $\sigma$-$\pi$ with $-40<\phi<40$ or
$140<\phi<220$; $\pi$-$\sigma$ with $40\lesssim\phi\lesssim140$ or
$220\lesssim\phi\lesssim320$. To obtain the phase for other $\phi$
angles one needs to shift these by 180 degrees.}
\label{fig:600phase}
\end{figure}

We notice that the resonant intensity goes to zero at about 1135eV
which is a strong indication that the phase of the mixing changes
sign. In order to study this we show in Fig.\ref{fig:600phase}
which shows a strong energy dependence and a change of sign at
about 18 eV above the edge and again at about 30 eV above the
edge. In fact we predict another resonance at higher energies and
this should be looked for in the future. In addition to the
prediction of the second resonance we also predict that the phase
of the scattering changes by $3\pi$. This change of phase should
change the interference with the omweg scattering
\cite{Renninger37} dramatically and also this should be studied as
a function of energy.

In agreement with the paper by Templeton and Templeton
\cite{Templeton94} our calculations show that  only the
dipole-quadrupole term  contributes to the resonant intensity in
the crystal structure of diamond. First of all, the pure dipole as
well as quadrupole term in Eq.\eqref{eq:basic} is essentially
proportional to local partial $p$ or $d$ density of states
projected on a particular site. Due to the fact that there is no
difference in the correspondent partial densities of states of the
8 germanium atoms in the unit cell and the reverse of signs in
structure factor the intensity of resonant scattering in
dipole-dipole or quadrupole-quadrupole channels is zero. Secondly,
as  already mentioned, the mixed dipole-quadrupole term involving
transition from 1$s$ to 4$p$ and from 4$s$ to 1$s$ orbitals of
germanium also does not give any contribution to the resonant
scattering because neither of these are quadrupole allowed. This
means that sp3 hybridization will not contribute to this
scattering.

The net tensor has then  only two significant contributions which,
in for example the  $p$-$d$ dipole-quadrupole transition has a
symmetry:
\begin{displaymath}
\begin{pmatrix}
0 & 0 & 0 & C & 0\\
C & 0 & 0 & 0 & 0\\
0 & C & 0 & 0 & 0
\end{pmatrix}
\end{displaymath}
Here $C$ is some constant for each particular energy and
independent on azimuthal angle $\phi$. The rows and columns of
this matrix are numerated according to quantum number order ($m =
-l \ldots l$). The reverse transition has exactly the same
symmetry and sign but both of them have opposite sign to the
matrix elements related to the second type of germanium atoms in
the unit cell. The latter is a direct consequence of wave function
symmetry caused by the particular  mixing of  $p$ and $d$ orbitals
in the diamond crystal structure.

\begin{figure}
\centering
\includegraphics[clip=true,width=0.36\textwidth]{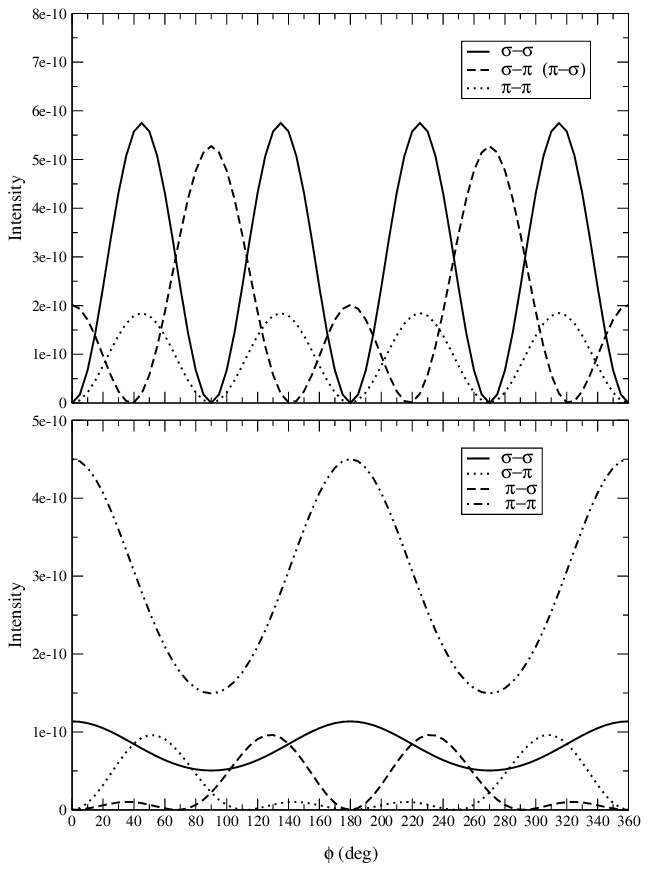}
\caption{The azimuthal angle dependence of resonance peak at
energy 11109.2  eV. The upper panel shows 600 reflection for
scattering angle $\theta=36.35^\circ$. The lower panel shows 222
reflection at $\theta=20.01^\circ$.} \label{fig:bpik}
\end{figure}

Finally we present the calculated azimuthal angle dependence of
the 600 and 222 reflections at K-edge. Fig.\ref{fig:bpik} shows
that the intensity of the 600 reflection follows a
$(sin\phi\,cos\phi)^2$ behaviour in $\sigma$-$\sigma$ and
$\pi$-$\pi$ channels with $\phi$ equal to zero when the (001) axis
is in the scattering plane and the incident beam is approximately
parallel to this axis Fig.\ref{fig:geom}.For the 222 reflection it
changes to more like  $cos^2\phi$ behaviour whereas the period in
$\sigma$-$\pi$ ($\pi$-$\sigma$) channel is changing from $\pi$ to
$2\pi$.

For the 600 reflection we also calculate the absolute value of the
structure factor following the paper by Colella and Merlini
\cite{Colella66} where we choose the polarization factor and the
Debye-Waller factors to be 1. To compare our calculations with
experiment we define the integrated intensity $I$ as $\sum_s
I^s_{\pi-\pi}+I^s_{\sigma-\sigma}$ were $s$ is a spin index. In
that case for an azimuthal angle of 25 degrees the integrated
intensity is calculated to be $1.656\cdot10^{-10}$ and the
structure factor 0.44 which is in excellent agreement with 0.475
determined from the experimental data by Colella \cite{Colella01}.

In conclusion we have demonstrated that the quadrupolar-dipolar
resonant scattering in Ge will indeed yield a 600 scattering
intensity and that this is a direct measure of the $p$-$d$
hybridization as a function of energy in the conduction band. Band
theory does an excellent job of explaining both the magnitude and
the energy dependence of the resonance. In addition we have made a
number of predictions regarding more resonances at somewhat higher
energies and we have predicted the phase. Future experiments
related to these predictions  should provide an even more critical
test of the theory. Since such $p$-$d$ mixing is only allowed if
the atom is not at an inversion center this mechanism could also
be important for all such materials and may explain part of the
structure seen at resonance for the transition metal oxides in
general and V$_2$O$_3$ in particular.

This investigation was supported  by the Netherlands Organization
for Fundamental Research on Matter (FOM) with financial support by
the Netherlands Organization for the Advance of Pure Science (NWO)
and  by Russian Foundation for Fundamental Investigations (RFFI
grants RFFI-01-02-17063).


\begin{thebibliography}{99}
\bibitem{Colella01} T.L.~Lee, R.~Felici, K.~Hirano, B.~Cowie, J.~Zegenhagen and
    R.~Colella,
    unpublished.
\bibitem{Murakami98} Y.~Murakami, J.P.~Hill, D.~Gibbs, M.~Blume,
    I.~Koyama, M.~Tanaka, H.~Kawata, T.~Arima, Y.~Tokura, K.~Hirota, and
    Y.~Endoh, Phys. Rev. Lett. {\bf 81}, 582 (1998).
\bibitem{Ishihara98} S.~Ishihara, and S.~Maekawa,
    Phys. Rev. Lett. {\bf 80}, 3799 (1998).
\bibitem{Elfimov99} I.S.~Elfimov, V.I.~Anisimov, and
    G.A.~Sawatzky, Phys. Rev. Lett. {\bf 82}, 4264 (1999).
\bibitem{Benfatto99} Maurizio~Benfatto, Yves~Joly, and Calogero~R.~Natoli,
    Phys. Rev. Lett. {\bf 83}, 636 (1999).
\bibitem{Paolasini99} L. Paolasini, C. Vettier, F. de Bergevin,
    F. Yakhou, D. Mannix, A. Stunault, W. Neubeck, M. Altarelli,
    M. Fabrizio, P. A. Metcalf, and J. M. Honig,
    Phys. Rev. Lett. {\bf 82}, 4719 (1999).
\bibitem{Ezhov99} S.Yu.~Ezhov, V.I.~Anisimov, D.I.~Khomskii
    and G.A.~Sawatzky, Phys. Rev. Lett. {\bf 83}, 4136 (1999).
\bibitem{Mila00} F.~Mila, R.~Shiina, F.-C.~Zhang, A.~Joshi, M.~Ma,
    V.~Anisimov, and T.M.~Rice, Phys. Rev. Lett. {\bf 85}, 1714 (2000).
\bibitem{Templeton94} D.H.~Templeton and L.K.~Templeton,
    Phys. Rev. B {\bf 49}, 14~850 (1994).
\bibitem{Ewald36} P.P.~Evald and H.~H\"{o}nl,
    Ann. Phys. {\bf 25}, 281 (1936).
\bibitem{Colella66} R.Colella and A.Merlini,
    Phys. Stat. Sol. {\bf 18}, 157 (1966).
\bibitem{Colella74} R.Colella,
    Acta Cryst. A {\bf 30}, 413 (1974).
\bibitem{lmto} O.~K. Andersen, Phys. Rev. {\bf B12}, 3060 (1975).
\bibitem{Renninger37} M. Renninger,
    Z. Phys. {\bf 106}, 141 (1937).
\end{thebibliography}
\end{document}